\documentclass[preprint]{emulateapj}

\newcommand{\EQ}{\begin{equation}}
\newcommand{\EN}{\end{equation}}

\newcommand{\Fig}[1]{Figure~\ref{#1}}

\newcommand{\pc}{{\rm\,pc}}

\shorttitle{Angular Momentum Transport in Extended Galactic Disks}
\shortauthors{McNally, Wadsley, Couchman}

\begin{document}

\title{Self-Gravity and Angular Momentum Transport in Extended Galactic Disks}

\author{C.P.\ McNally\altaffilmark{1,2}}
\email{colinm@astro.columbia.edu}
\author{J.\ Wadsley}
\email{wadsley@mcmaster.ca}
\author{H.M.P.\ Couchman}
\affil{
Department of Physics \& Astronomy,
McMaster University,
1280 Main St. W,
Hamilton, ON,
L8S~4M1,
Canada}
\email{couchman@mcmaster.ca}

\altaffiltext{1}{Current address: Department of Astronomy, Columbia University, 550 West 120th Street, New York, NY, USA 10027}
\altaffiltext{2}{Corresponding author}
\begin{abstract}
We demonstrate a significant difference in the angular momentum transport 
properties of galactic disks between regions in which the interstellar 
medium is  single phase or two phase.
Our study is motivated by observations of \ion{H}{1} in extended galactic disks which 
indicate velocity dispersions of nonthermal origin, suggesting that 
turbulence in the gas may be contributing significantly
to the observed dispersion.
To address this, we have implemented a
shearing-box framework within the FLASH code.
The new code was used to perform local simulations of galactic disks that 
incorporate differential rotation, self-gravity,
vertical stratification, hydrodynamics and cooling.
These simulations explore plausible mechanisms for driving
turbulent motions via the thermal and self-gravitational instabilities
coupling to differential rotation.
Where a two-phase medium develops, gravitational angular momentum 
transporting stresses are
much greater, creating a possible mechanism for transferring energy from 
galactic rotation to turbulence.
In simulations where the disk conditions do not trigger the formation of a 
two-phase medium,
 it is found
that perturbations to the flow damp without leading to a sustained mechanism 
for driving turbulence.
The differing angular momentum transport properties of the single- and 
two-phase regimes
of the disk suggest that a significant, dynamically motivated division can be 
drawn between the two,
even when this division occurs far outside the star formation cutoff in a 
galactic disk.
\end{abstract}

\section{Introduction}
\label{secintroduction}
The magnitude of the \ion{H}{1} velocity dispersion in the outer regions of galactic 
disks
is observed to be surprisingly consistent, 
about $7 \pm 1{\rm\,km\,s^{-1}}$ (\citeauthor{2007ApJ...662L..19S} \citeyear{2007ApJ...662L..19S}; 
particular examples 
are  
DDO 154, \citeauthor{1990ApJ...352..522D}~\citeyear{1990ApJ...352..522D};
ESO 215-G?009, \citeauthor{2004AJ....128.1152W}~\citeyear{2004AJ....128.1152W}; and those 
discussed by \citeauthor{2006ApJ...638..797D}~\citeyear{2006ApJ...638..797D} and references therein).
Though supernovae are accepted to be the main energy source for interstellar (ISM) turbulence where
star formation occurs (see review by \citeauthor{2004RvMP...76..125M} \citeyear{2004RvMP...76..125M}) 
the driving mechanism for turbulence where
this is not the case is unclear.
Theoretical arguments have been made by
\citet{1999ApJ...511..660S} and
\citet{2007ApJ...663..183P} that this turbulence is magnetorotational instability (MRI) turbulence, and by 
\citet{2007ApJ...662L..19S} that it is driven through the infall of gas clouds.

Typically, the stellar disk does not stop immediately at a well-defined radius; 
most commonly the rate at which the brightness profile
declines switches to a much steeper exponential than in the inner disk.
This double exponential profile has been characterized by
\citet{2007arXiv0706.3830P}.
Theories  predicting a sharp cutoff to star formation,
such as the $Q$ criteria of \citet{1989ApJ...344..685K} and 
the cold-phase criteria of \citet{1994ApJ...435L.121E}
and \cite{2004ApJ...609..667S} do not naturally explain this profile.
However, the theory of \citet{2006ApJ...636..712E}, which includes a second regime of 
turbulence-triggered star formation outside the cutoff radius, may 
produce such double exponential profiles.
Additionally, \citet{2006ApJ...645..209D} suggest that an 
initial single-exponential brightness profile could evolve into
a double-exponential one through secular effects.

Low-level star formation in the extended disk would require the cold phase to exist in at least some regions,
indeed
\citet{2006AJ....131..363D}
observed both a warm and a cold phase outside the observed limit of star formation in NGC~6822.

Self-gravity  in conjunction with differential rotation has been examined as a 
mechanism for driving turbulence by 
\citet{1999ApJ...516L..13W,
2001ApJ...547..172W,
2007ApJ...660..276W}
and
\citet{2002ApJ...577..197W}
with global two-dimensional and three-dimensional models of a galactic disk 
including self-gravity and cooling. 
In order to release energy with which to drive turbulence, 
self-gravitational interactions must lead to angular momentum transport.
Hence it is important to analyze the criteria under which self-gravitational angular momentum 
transport can occur.

Following \citet{1972MNRAS.157....1L} and
\citet{1999ApJ...511..660S}
we introduce the mass-weighted average angular momentum
transporting stress $\langle T_{xy}\rangle$.
The $xy$ subscript results from this being evaluated in a 
local set of Cartesian coordinates around one point in  the disk.
In this frame, $x$ is the radial coordinate and 
$y$ is the azimuthal coordinate.
The gas perturbed velocity is denoted by $\mathbf{u}$ and the
gravitational velocity is defined as
$\mathbf{u}_G \equiv {\nabla \Phi}/{\sqrt{4\pi G \rho}}$
following \citet{1999ApJ...511..660S}.
Then the $xy$ component of the stress, which transports angular momentum, is
\begin{equation}
\langle T_{xy} \rangle \equiv \langle \rho u_x u_y+ \rho u_{Gx} u_{Gy}\rangle.
\end{equation}
The first term in this expression is the hydrodynamic stress,
 or $T^\mathrm{hydro}_{xy}$,
and the second is the gravitational stress,
 or $T^\mathrm{grav}_{xy}$.
The mass-weighted average values of these, $\langle T_{\mathrm{}xy}\rangle$,
will be used to detect whether a disk has the ability to drive turbulence
through the coupling of differential rotation to local motions.

\citet{2007ApJ...660..276W} have performed global simulations
of self-gravitationally driven turbulence in a galactic disk.
In comparison to the initial conditions used in this work (see Section \ref{secfixedmass})
the initial conditions of \cite*{2007ApJ...660..276W}
lie far into the thermally unstable regime.
Those simulations are not accompanied by a direct analysis of the role of angular 
momentum transport in driving the turbulence.
This paper carefully examines 
the evolution of angular momentum transport.
\citet{2007ApJ...663..183P} have performed local simulations of MRI in the context
of a galactic disk with a two-phase medium, but do not examine self-gravity.
In this paper, we present the results of simulations showing the 
difference in self-gravitational angular momentum transport properties of single-phase 
and two-phase galactic-type disks.
We also demonstrate the onset of this mode of angular momentum transport 
when a single-phase accreting disk gains sufficient 
surface density to form a two-phase medium. 
This was done by a series of models of local disk patches, with either fixed 
mass or continuous accretion.
The possible regimes of behavior for a local patch of  galactic disk are then discussed.

\section{Methods}
We began with the standard FLASH 2.3r1 code \citep{2000ApJS..131..273F}
which, among other features, included support for parallel 
adaptive mesh hydrodynamics and self-gravity in periodic boundary conditions.
We implemented support for shearing-box boundary conditions and fictitious forces,
shearing-box self-gravity, and modified the hydrodynamics for the shearing frame.
The background shearing flow was split from the hydrodynamics, 
in a manner similar to 
\citet{2000A&AS..141..165M} and \citet{2001ApJ...553..174G}.
The evolution of linear shearing waves has been tested, as in
\citet{2005ApJ...635..149J}
and \citet{2006ApJ...653..513S}.

To handle the combination of both self-gravity (Poisson's equation)
and the shearing frame (Hill's equations; \citeauthor{hill1878}, \citeyear{hill1878})
three choices are made.
First, the vertical background force term, originally $\Omega^2z$, was 
transitioned to $0$ in the range
$z_c<z<2z_c$, where $z_c$ is a parameter, 
 by means of a cubic interpolation.
The vertical external gravity was hence zero at the vertical boundaries.
Second, the vertical fluid boundary is set to either zero-gradient (outflow)
or a fixed-value (inflow).
Third, self-gravity is solved with a periodic boundary condition in the vertical direction,
and the simulation box kept tall and thin
so that the combination of the self-gravity and the vertical background force 
produces a boundary condition essentially equivalent to an isolated boundary condition for the disk.

The shearing box is specified by two parameters, 
the angular frequency $\Omega$ and the radial rate of change of this quantity, 
$q\equiv -d\ln \Omega /d\ln R$.
To simulate an outer galactic disk, these are chosen 
as $\Omega = 8.9\times10^{-16}{\rm\,s^{-1}}$
and $q=1$.

As the shearing box is a local model,
global self-gravitational phenomena such as spiral arms and bars cannot be captured.
The shearing box also forces the angular momentum transport to be purely local, 
i.e.\ within the scale of the box \citep{1999ApJ...521..650B}.
In the context of this work, this is an advantage, as the mechanisms produced 
will be local and minimal in nature, not relying on the large-scale structure of the galaxy,
and hence are, perhaps, more generally applicable.

\subsection{Cooling}
We investigate the significant difference in behavior 
between single- and two-phase disks.
The formation of a two-phase disk is dependent on the radiative cooling used.
To include the local effects of various optically thin limit
cooling and heating mechanisms a net heat loss function is defined as 
${\cal L}(T)\equiv\rho\Lambda(T)-\Gamma$,
where $\Gamma$ is a constant following  \cite*{2007ApJ...654..945B} and others.
We use  a heating parameter $\Gamma = 0.015\,{\rm erg}\,\mathrm{g}^{-1}\,\mathrm{s}^{-1}$ as in
 \citet{2007ApJ...654..945B} and  \citet{2007ApJ...663..183P}.
The form of $\Lambda(T)$ is a piecewise power-law fit from  \citet{2007ApJ...654..945B}.
The resulting heating equilibrium curve is shown in \Fig{col3rhop2}.
For calculating temperature, the mean particle mass is $\mu=0.62$ and 
a $\gamma=5/3$ ideal gas equation of state is used.
To ensure that the Jeans length is resolved, 
the Jeans length limiter is taken from \citet{2007ApJ...660..276W}. 
This is designed to prevent artificial fragmentation by 
enforcing a minimum value of the Jeans length \citep{1997ApJ...489L.179T}.
The Jeans limiter artificially halts collapse where is can still be resolved without artificial fragmentation.
Runs were stopped when the computational expense became prohibitive.
Whether the collapse stops in real disks, and how, cannot be addressed by these models. 

\section{Results}

\subsection{Fixed Mass Disk}
\label{secfixedmass}

The important parameter in fixed-mass disk models is where the phase of the
mid-plane lies in relation to the critical curve for thermal instability.  
Disks in which the densest parts have perturbations that are sufficiently
strong to trigger thermal instability, form a two-phase medium.
Seven models with disks of fixed mass and cooling were run.
 The models had parameters such that the mid-plane conditions varied around the heating-cooling equilibrium curve
on the $\rho$-$P$ phase diagram. These models provided characterizations of the possible
 two-phase and single-phase disks.
The  parameters of these runs were
a box $1.2$ kpc $\times$ $1.2$ kpc $\times$ $4.8$ kpc,
with no minimum level of refinement.
The standard resolution run used four levels of refinement
at the mid-plane or a mid-plane resolution of $128^2$ and $9.375\pc$.
The initial condition was a hydrostatic equilibrium isothermal disk, with 1\% 
random density perturbations added to each cell.
Runs with surface density $\Sigma=6{\,\rm M_\odot\pc^{-2}}$ 
and initial temperature $2500\,{\rm K}$, the standard mid-plane resolution, and double the standard resolution
developed a two-phase medium.
An additional variation with 
$\Sigma=15{\,\rm M_\odot\pc^{-2}}$ and initial temperature $10,000\,{\rm K}$ also developed this two-phase medium.
The unstable filaments connecting the lumps are gradually accreted into clouds
and an unstable layer exists as a sheath around the cool clouds.
A $Q$-stable disk of thermally stable gas remains extending far above the clouds.
The phase diagram of this two-phase medium consists of points lying on the warm 
thermal equilibrium line, a scatter of points proceeding across
to the cold equilibrium line below the peak warm equilibrium pressure,
points lying on the cold equilibrium line, and a small number of 
points lying on the Jeans limiter line.
This is similar to the phase diagram for the late time of the accreting disk 
shown in \Fig{col3rhop2}.
When the disk is single-phase, it evolves toward a 
smooth shearing flow without significant angular momentum transporting stress. 
The four remaining runs evolved to this state.
Additionally, Toomre-stable isothermal gas disks (without cooling) also behaved in this manner.
 
Even in the two-phase disk runs, the total mass-weighted value of the Toomre $Q$ parameter 
\citep{1964ApJ...139.1217T,1965MNRAS.130..125G}  stays well in the range of stability.
For example, $Q>4.5$ for all times in the first two-phase run described,
 even though the cool phase by itself is Toomre-unstable. 
The  self-gravitational stresses of two-phase fixed mass disks are essentially the same as the two-phase part of the accreting disk to be discussed in Section \ref{secaccreting}.

\subsection{Accreting Disk}
\label{secaccreting}

\begin{figure}
\plotone{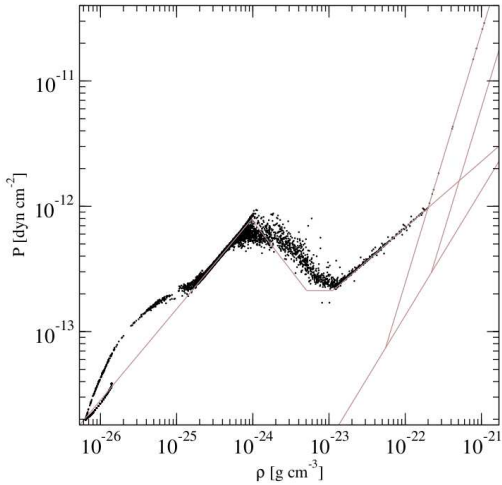}
\caption{Run A1 gas phase distribution at $t= 2.73\times 10^{16}{\rm\,s}$ with equilibrium curve and temperature limits}
\label{col3rhop2}
\end{figure}

\begin{figure}
\plotone{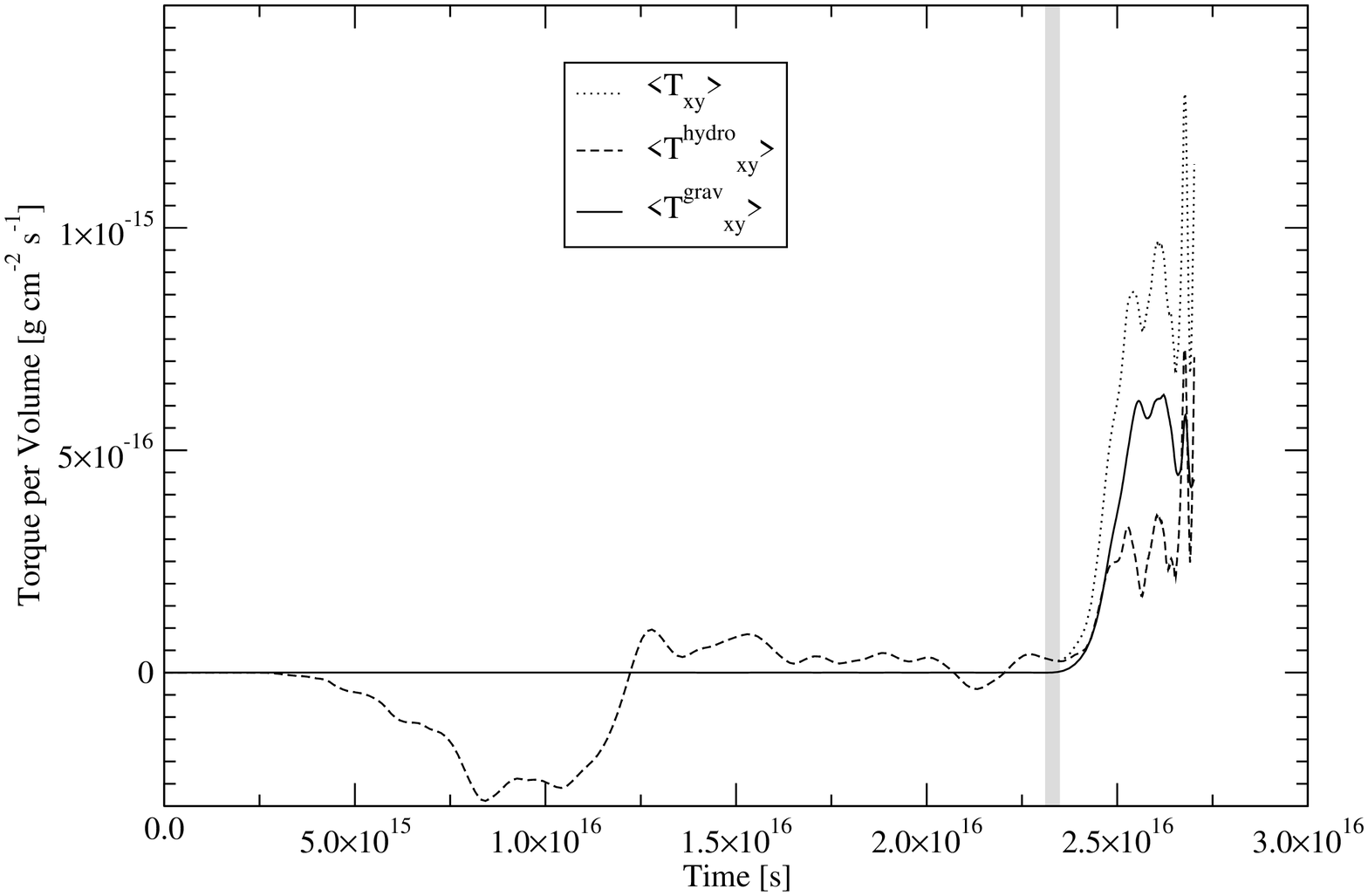}
\caption{Run A1 stress -- showing the transition from stable to unstable evolution as the cool phase develops, the gray vertical stripe marks the onset of the formation of the cool phase, the jump in $\langle T^\mathrm{grav}_{xy}\rangle$ is approximately a factor of $10^4$ }
\label{col3stressfig}
\end{figure}

Run A1 starts with the grid as in the standard resolution fixed-mass disk in the previous section, but with an initial condition of 
uniform very low density ($10^{-28}{\rm\ g\,cm^{-3}}$) gas.
Inflow $z$-direction boundaries 
were specified
with the inflow rate set as $1.146\times10^{-8}{\rm\ M_\odot\,pc^{-2}\,yr}$.
Along the $16$ $\times$  $16$ grid of cells at the $z$-boundary, the density is
$1\%$ greater on odd-numbered rows.
This provides a minimal but nonangular momentum transporting
perturbation to the flow, 
 so that the imposed perturbations 
will be larger than those arising from purely numerical effects.
With these boundary conditions, the disk built up and
eventually reached the critical
temperature/pressure for thermal instability and formed a cool phase. 
The inflow rate was chosen to be slow so that the cool phase formation is not triggered
directly by the inflow pressure.

The disk starts at very low surface density and slowly builds.
As the infalling gas from either boundary collides with the disk it drives
turbulence in the disk as it builds.
This is however purely a function of the inflow boundary conditions.
As the disk surface density grows, so does the mid-plane pressure,
until eventually it is sufficient to trigger thermal instability, 
and the cool phase readily forms.
Turbulence driven by the infalling gas creates density perturbations 
which the formation of the cool phase amplifies.
As these cool lumps form and rapidly accrete gas, the gravitational 
interactions between them lead to vastly increased 
angular momentum transporting gravitational stress, as shown in
\Fig{col3stressfig}.
When the gravitational stress amplifies as a result of the phase transition the
change is dramatic: in the case of run A1 it is $4$ orders of magnitude.
After the cool phase forms, the phase diagram is like that displayed in \Fig{col3rhop2}.
Gas is introduced at a point near the heating-cooling equilibrium, initially
expands and cools as it falls in, then recompresses and heats,
 rising above the equilibrium line toward higher densities.
There is then a layer of gas lying along the equilibrium line, until the critical pressure
for the phase change is reached.
A further scatter of points on the phase plot shows the layer of cells 
around the cool clouds where gas is unstable and cooling onto the clouds.
Finally, points denoting the cool clouds are distributed along the cool equilibrium line.

The run A2 repeats the same setup as run A1, but with half the mass inflow rate.
A2 has the same pattern of rapidly growing stresses, with the
transition at approximately the same surface density, 
demonstrating the critical importance of surface density rather than sensitivity to the 
imposed inflow rate.
Between runs A2 and A1 the pre-transition hydrodynamic 
stresses vary in a  different manner, 
demonstrating that they are indeed a function of the imposed boundary conditions.

\section{Discussion And Conclusions} 
Two primary conclusions can be drawn from this work.
First, thermal instability can trigger self-gravitational angular momentum transport.
Second, the development of the cool phase leads to rapidly growing local self-gravitational angular momentum transporting stress.

The thermally stable cooling disks of fixed mass
decayed toward a smooth flow state. 
Similarly,
before the onset of thermal instability, the accreting
disk models did not have significant self-gravitational stress.
However, the combined results of the fixed-mass disk with cooling and the accreting disk
simulations show that thermal instability
and the formation of a two-phase medium can provide a mechanism for 
producing enhanced angular momentum transport by
self-gravitational stresses.
This will have two consequences:
 the galaxy's rotational energy is converted into
local motions giving rise to velocity dispersion
in the gas; and it enhances the rate at which mass,
 primarily in the form of cool gas clouds, is transported inward 
enhancing the surface density of more central regions of the galactic disk.
Increased surface density may in turn lead to star formation.

If the observed roughly constant velocity dispersion in the outer galactic disk
is to be driven by a local self-gravitational
mechanism, the simulations in this work suggest that the cool
phase must exist everywhere in the disk where the velocity dispersion is seen.
However, as the surface density of the disk drops, 
it will be increasingly difficult to achieve the critical
pressure for the existence of the cool phase at the disk 
mid-plane.
Detailed observations, extending the work of  
\citet{2006AJ....131..363D},
might settle this by directly detecting a cool phase in 
extended, low-surface-density galactic disks.

\begin{figure}
\plotone{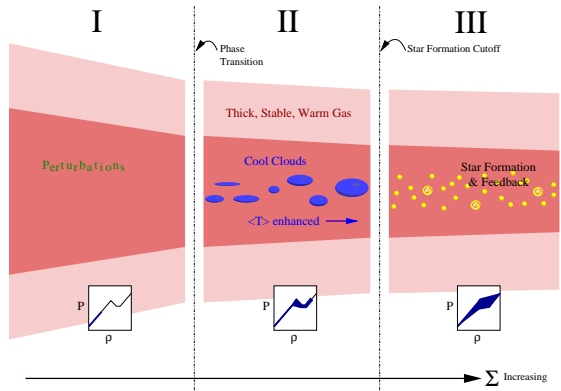}
\caption{Three regimes of a galactic disk, the $P-\rho$ insets show which parts of the phase diagram are populated around the cooling-heating equilibrium curve}
\label{galpicfig}
\end{figure}

Finally, a picture of the phase structure and dynamic structure
of the outer regions of a galactic disk can be drawn.
Three regimes, as shown in \Fig{galpicfig}, can be identified,
with the divisions between them drawn by the cutoff in star formation and the
gas single-phase to two-phase transition.
\renewcommand{\labelenumi}{\Roman{enumi}}
\begin{enumerate}

\item A region of single phase gas, closely following the heating-cooling equilibrium curve on the phase diagram,
with $Q>1$, and with pressure too low to support a two-phase medium. Includes small perturbations. The total disk is 
Toomre-stable.
\item A region where pressure is sufficient for cool clouds to form from existing small perturbations,
forming a two-phase medium, enhanced angular momentum transporting stresses result. The total disk is Toomre-stable.
\item The stellar disk, where the conditions for star formation are widely satisfied, and feedback (supernovae
and outflows) drives turbulence. The total disk has Toomre $Q$ satisfying a \cite{1989ApJ...344..685K} relation.
\end{enumerate}
Regions {\sc I} and {\sc II} are simple to form - thick warm gas disks are physically simple,
and the stellar disk is well known to exist.
There are already a number of elements pointing to the nature of region {\sc II}.
\citet{2007ApJ...663..183P} point out that the  \citet{2004ApJ...609..667S}
hypothesis, that the formation of the cool phase necessarily implies star formation,
 fails based on current observations showing that the phase
change can occur outside the cutoff of star formation.
That a region where turbulence must be driven by nonstellar processes while the cool phase 
exists has also been argued by \citet{2007IAUS..237...70O}.
In the same work, they also find that magnetohydrodynamic effects in a region with cool clouds,
such as in region {\sc II}, may suppress star formation.
The presence of star formation sets the division between regimes
{\sc II} and {\sc III}.
This cutoff may not be sharp
and it could be expected that locally star formation
may happen in the cool phase of region {\sc II}.
The presence of stars, and their feedback (particularly supernovae,
see  \citeauthor{2004RvMP...76..125M}, \citeyear{2004RvMP...76..125M})
 changes the dynamics and phase structure of
gas from {\sc II} to {\sc III}.
Between regimes {\sc I} and {\sc II} the dividing line is drawn on
the existence of the cool phase.
In terms of the importance of local self-gravity in galactic disks,
this work has shown that there is a significant energetic divide between 
regimes {\sc I} and {\sc II}.

The self-gravity and hydrodynamic stresses presented here in media with nonzero shear 
can be calculated and can be of similar magnitudes in nonshearing cases.
A version of run A1 without shear was preformed to demonstrate this. 
The definition of the stresses does not depend explicitly on the existence of nonzero shear. 
However, without shear, these stresses are not significant as they cannot extract energy 
from the background shear or transport angular momentum without a nonzero shear to couple to.

In regime {\sc II} simulations in this work, the mass weighted average power 
per unit volume from self-gravitational stresses is typically on the order of 
$10^{-29}\mathrm{\ erg\ cm^{-3}\ s^{-1}}$, a significant energy input in 
such a context \citep{2004RvMP...76..125M}. We have also demonstrated that 
the presence of self-gravity leads to larger velocity fluctuations as 
follows. In a fixed-mass disk, we evolved a  $\Sigma=6{\,\rm 
M_\odot\pc^{-2}}$ disk with an initial temperature of $1250\,{\rm K}$ until a 
two-phase medium had developed. At this time the mass wighted RMS velocity of 
gas below $1000\ \mathrm{K}$ was $1.3\,\mathrm{km\ s^{-1}}$. This state was 
then used as input for two further simulations: one with the normal 
self-gravity, and the second  with a fixed vertical gravity that was obtained 
by horizontally averaging the vertical self-gravity forces in the input 
state.  When it became computationally infeasible to evolve the simulations 
further, the self-gravity case had a mass wighted RMS velocity of gas below 
$1000\ \mathrm{K}$ that had grown to $2.6\,\mathrm{km\ s^{-1}}$ and was 
climbing, while the same quantity in the fixed-gravity case had fallen to 
$0.8\,\mathrm{km\ s^{-1}}$ and was declining.

The extent to which extended \ion{H}{1} disks have regions of type {\sc II}
must be determined by observations detecting or showing the absence 
of a cool component. 
One existing example is the observations by \citet{2006AJ....131..363D}  
of NGC~6822, which suggest that part of this disk may be characterized as regime {\sc II}.
Even in the relatively small box used in Section \ref{secfixedmass}, 
the mass wighted RMS velocity of gas below $1000\,\mathrm{K}$ 
is $2\,\mathrm{km\ s^{-1}}$. This is in the same ballpark as the observed \ion{H}{1} velocity dispersions.

\acknowledgments
The software used in this work was in part developed by the DOE-supported ASE / Alliance Center
for Astrophysical Thermonuclear Flashes at the University of Chicago.
This work was made possible by the facilities of  
{\sc SHARCNET}. 
C.P.M.\ was partially supported by NSF CDI grant AST08-35734.
H.M.P.C.\ acknowledges the support of NSERC and CIfAR.  
J.W.\ acknowledges the support of NSERC.


\end{document}